\documentclass[12pt]{iopart}
\usepackage{graphicx}
\usepackage{cite}

\newcommand{\bvec}[1]{\mbox{\boldmath $#1$}}

\begin{document}

\title{Experimental Verification of a Spin-Interference Device Action}

\author{Y Iwasaki, Y Hashimoto, T Nakamura and S Katsumoto}

\address{Institute for Solid State Physics, University of Tokyo, 5-1-5 Kashiwanoha, Kashiwa, Chiba 277-8581, Japan}
\ead{you.iwasaki@issp.u-tokyo.ac.jp}
\vspace{10pt}
\begin{indented}
\item[]July 2016
\end{indented}

\begin{abstract}
We report the detection of spin interference signal in an Aharonov-Bohm type interferometer with quantum dots on the conduction paths. 
We have found that resonators like quantum dots can work as efficient spin rotators. The interference signal appears only when spin-polarized electrons are injected into the device. The interference pattern in the gate voltage-magnetic field plane is checker board like, ensuring the modulation of spin wavefunction's phase as well as the orbital phase. 
\end{abstract}

%
%
%
%
%

\section{Introduction}

In spite of quantum mechanical nature of the spin freedom in electrons, it has been mostly regarded as a classical variable in traditional spintronics. That is, a spin is treated as a single (classical) bit, which carries information  up or down\cite{wolf}.
Actually a spin can work as a quantum bit (qubit) or even as a ``flying qubit", which is expressed as a linear combination of the two (up and down) eigenstates.
This fact enables us to compose quantum spintronics devices, which would gain novel functions for information processing and greatly widen the field of spintronics.
The quantum nature of spins prominently appears in the interference\cite{nitta1, miller}.
A device which utilizes such spin interference of electrons traversing through it was 
proposed by Aharony {\it et al.}\cite{aharony}. They showed that a diamond like simple interference device can serve as an efficient and precise spin rotator and an analyzer.
This device requires spin-orbit interaction (SOI) to rotate the spin of electrons traversing over the diamond structure  and the Aharonov-Bohm (AB) phase gained from magnetic flux.

Here we report experimental verification of such device action in an AB interferometer with quantum dots on the two transmission paths.
Since the spin rotation was achieved by quantum dot resonators in our device, its magnitude depends on initial spin polarization. This effect enables us to confirm the operation of the spin rotators by injecting electrons from a quantum point contact (QPC), because the interference signal appears only when spin-polarized electrons were injected and QPC can tune its polarization.

\section{Methods}

\begin{figure}[b]
\includegraphics[width=\linewidth]{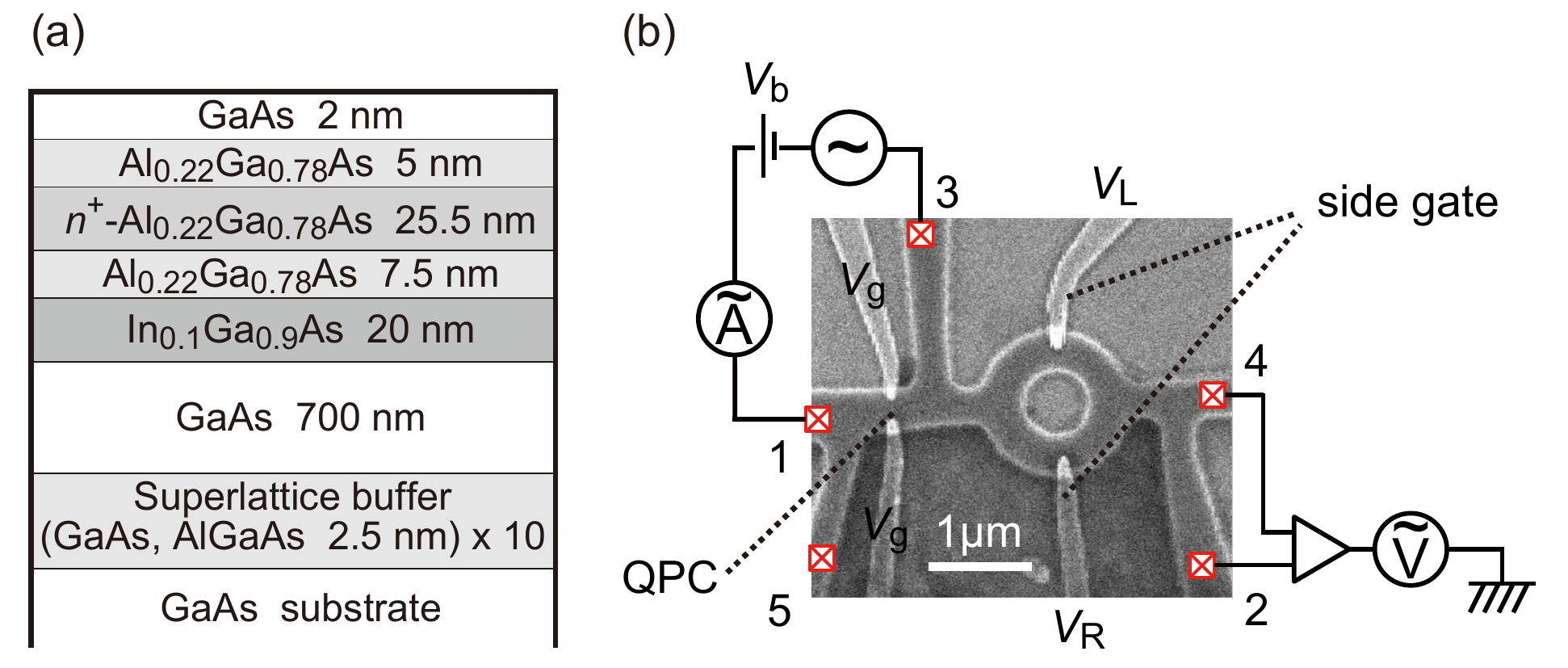}
\caption{(a) Cross sectional view of the layered structure. (b) Scanning electron micrograph of the sample with the terminal numbers and schematic non-local resistance measurement wiring.}
\label{f1}
\end{figure}
Figure \ref{f1}(a) shows the cross-sectional view of layered structure grown with ordinary molecular beam epitaxy onto a (001) GaAs substrate.
A pseudo-morphic In$_{0.1}$Ga$_{0.9}$As quantum well is placed next to an Al$_{0.22}$Ga$_{0.78}$As spacer.
It is well known that structural asymmetry and narrowness in the energy gap bring about strong SOI\cite{zawatzky, yoh} and the present structure is reported to have a comparatively strong Rashba-type SOI\cite{kim2}.
The electron mobility $\mu$ = 65000~cm$^2$/Vs and the sheet carrier concentration $n = 1.1 \times 10^{12}~{\rm/cm^2}$ were obtained from the Hall and the Shubnikov-de Haas measurement at 4.2~K.
 An advantage of the present structure over InAs quantum well with stronger SOI's
 is availability of the conventional metallic split-gate technique for defining fine structures such as quantum point contacts (QPCs).
Figure \ref{f1}(b) is a scanning electron micrograph of the sample, which consists of a QPC (electron emitter) and an AB-type interferometer.
Each arm of the AB interferometer has a short side gate for the control of the conductance indicated as $V_{\rm L}$ and $V_{\rm R}$ in the figure. It will be shown later that these gates can pinch the paths and form quantum dots on them.
The electrodes and the gates are numbered as superposed on the figure for convenience of indication.

The specimen was cooled down to 0.15~K in a dilution fridge under a bias-cooling condition of the gate electrodes
for leakage free application of gate voltages\cite{ladriere}.
External magnetic field was applied perpendicular to the growth plane with a superconducting solenoid.
Two-wire and four-wire resistances were measured with conventional lock-in technique with frequencies lower than 80~Hz.

\begin{figure}
\includegraphics[width=\linewidth]{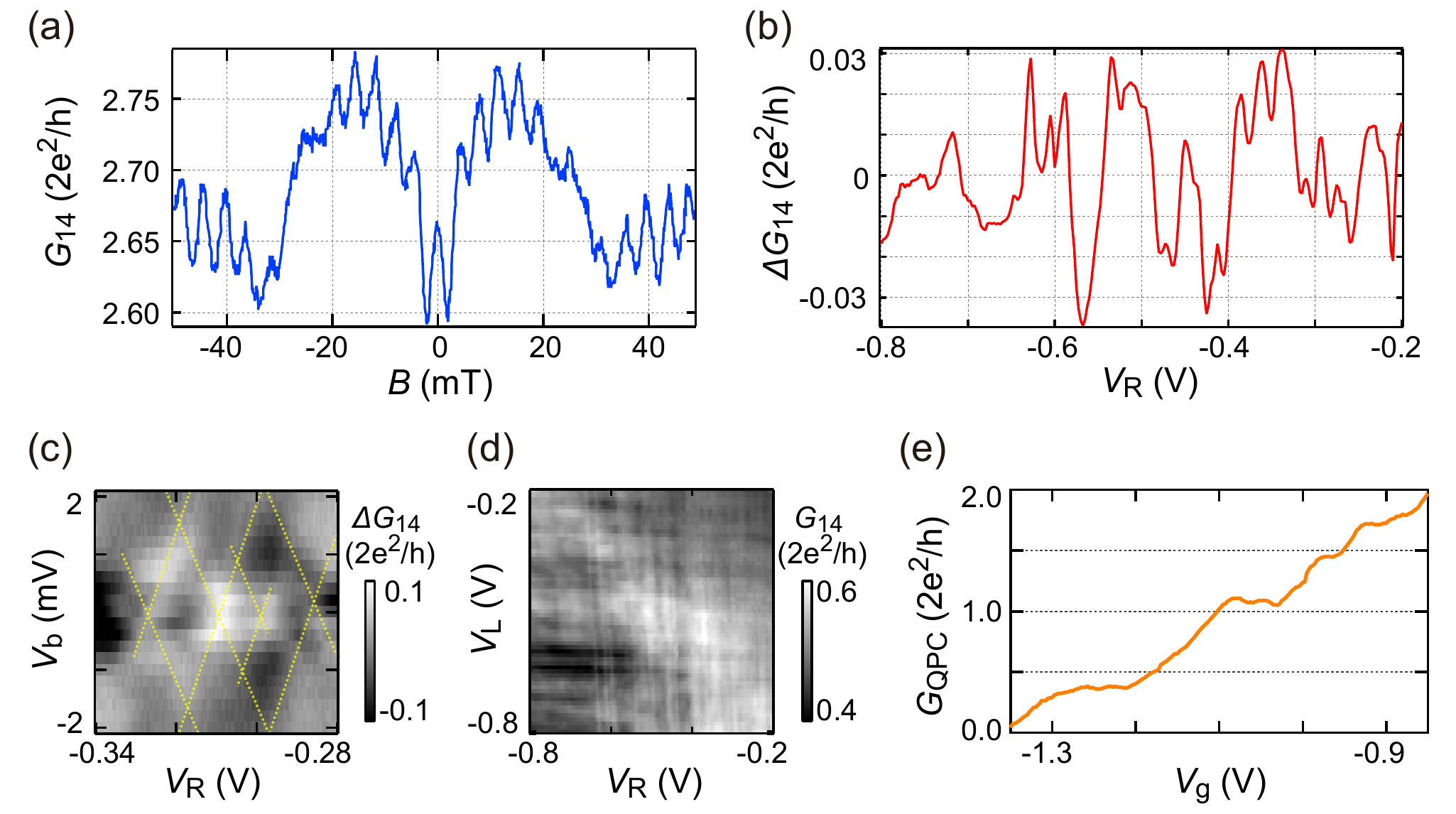}
\caption{(a) AB oscillation in two-terminal (local) conductance $G_{14}$.  (b) Conductance oscillation $\Delta G_{14}$ versus gate voltage $V_{\rm R}$ where $V_{\rm L}$ is kept to be $-0.79$~V. A linear background was subtracted.  (c) A gray scale plot of $\Delta G_{14}$ on $V_{\rm R}$. Coulomb diamond structures are indicated by yellow lines.
(d) Gray scale plot of the device conductance as a function of $V_{\rm R}$ and $V_{\rm L}$. Crossing white lines are the Coulomb peaks.
(e) QPC conductance $G_{\rm QPC}$ as a function of the gate voltage $V_{\rm g}$. The bias voltage $V_{\rm b}$ was $-1.8$ mV. Plateau structures are observed around 0.5, 1.0 and 1.5$G_{\rm q}$. The temperature for the measurements is 150~mK.}
\label{f2}
\end{figure}

First we need to check action of  each ``part" in the device.
Figure \ref{f2}(a) shows the two-terminal conductance $G_{14}$ (1 and 4 indicate the terminal numbers) as a function of the magnetic field for 
open-arm condition ($V_{\rm L}=V_{\rm R}=0$). A clear Aharonov-Bohm (AB) oscillation with the period for magnetic flux $\phi_0\equiv h/e$ (flux quantum) in the ring area
is observed to be symmetric to the zero-field due to the Onsager reciprocity\cite{yacoby}. 
The result certifies that the ring is working as a quantum interferometer.

Next we apply negative voltages to the two gates on arms of the ring.
The conductance $G_{14}$ decreases with the negative gate voltages and before pinch-off thresholds, aperiodic
oscillations versus the gate voltages appear as shown in Fig.2(b) for $V_{\rm R}$.
The oscillation can be viewed as an overlap of two Coulomb oscillations: one with narrow peaks and short periods coming from localized states weakly coupled to the electrodes; the other with broad peaks and wide periods from states strongly coupled to the electrodes (referred as {\it strongly-coupled states})\cite{aikawa}.
Actually as exhibited in Fig.2(c), the current-voltage characteristics form several sizes of Coulomb diamonds overlapping each other.
Hence we interpret the oscillation as formation of quantum dots  around the ends of gate electrodes, which phenomenon often happens, {\it e.g.},
around the pinch-off of QPC\cite{webb,ando}.
The irregularity of the oscillations indicates that the quantum confinement to the dots also contributes to the conductance spectra and each peak corresponds to a single electron level. The gate capacitance is estimated from the smallest peak interval as about 20~aF, which is reasonable from the geometric dimensions.
Figure \ref{f2}(d) shows the conductance in a gray scale as a function of the two gate voltages.
The oscillation peaks appear as white grid-crossing lines. 

Our last check is the conductance of the QPC against the gate voltage, which is displayed in Fig.2(e).
A clear plateau structure is observed at around a half of quantum conductance ($G_{\rm q}\equiv 2e^2/h$) in addition to the one around a full $G_{\rm q}$.
This is the sign that the SOI is strong enough to realize a spin filter on the half $G_{\rm q}$ plateau and a spin rotator on the full $G_{\rm q}$ plateau\cite{kohda, kim}.

We then proceed to look for the spin interference effect.
The simple crossings ({\it i.e.}, no avoided-crossing) in Fig.2(d) suggest that the coherent portion in the conductance is small and to extract such transport, we need to change the terminal configuration.
It is well known that in so called non-local configuration, not only the coherent portion is emphasized in the total resistance, but also the coherence itself is enhanced by blocking of the external voltage fluctuation\cite{kobayashi,seelig} naturally built in the non-local configurations.
Figure 1(b) schematically displays the terminal configuration for the resistance measurement.
The electric current flows between terminals 1 and 3 through the QPC. This causes a local nonequilibrium around the current path,which propagates coherently from the current path to the detector.

The four-terminal transport characteristics expected for a spin-interference device can be calculated in the Landauer-B\"uttiker formalism\cite{buettiker}.
Let $T_{ij}$ be the transmission coefficient between terminals $i$ and $j$, and $R_{ij,kl}$ be the resistance for current flow between $i$ and $j$, voltage between $k$ and $l$.
From an S-matrix analysis\cite{imry}, the transmission coefficients across the ring can be generally written as\cite{gefen,oreg,molnar}
\begin{equation}
T_{ij}=\frac{a_{ij}+b_{ij}\cos(\omega/2)\cos(\phi+\delta_{ij})}{1+c_{ij}f(\phi, \delta_{ij}, \omega)},
\label{eq_transmission_ij}
\end{equation}
where $\phi$, $\delta_{ij}$, $\omega$ are the phase shifts due to the AB effect, the electrostatic potential, and the spin rotation respectively.
In the numerator, coefficient $a_{ij}$ represents the imbalance between the two interference paths. The second term can be understood within the simplest AB approximation. That is, the probability amplitude at the interference node 
with perfect balance is
\begin{eqnarray*}
\left[\left(\cos\frac{\theta}{2}+\cos\frac{\theta+\omega}{2}\right)\langle \alpha|+
\left(\sin\frac{\theta}{2}+\sin\frac{\theta+\omega}{2}\right)\langle\beta|\right]
\langle\psi|&&(1+e^{-i\phi})\times({\rm c.c.})\\&&\propto \cos\frac{\omega}{2}\cos\phi,
\end{eqnarray*}
where $|\alpha\rangle$ and $|\beta\rangle$ are spin up and down eigenstates.
With addition of channel dependent kinetic phase gained from electrostatic potential $\delta_{ij}$, the second term of the numerator in eq.(\ref{eq_transmission_ij}) is reproduced.
$f(\phi, \delta_{ij}, \omega)$ in the denominator also consists of trigonometric functions
representing the effect of multiple circulation on the ring. This is smaller than 1 and can be negligible in the discussion of interference pattern.
As can be seen in Fig.\ref{f1}(b), the ring part is common for $T_{ij}$'s in (\ref{eq_transmission_ij}) and 
it is natural to assume the ratio $a_{ij}/b_{ij}\equiv a_0$ is common in (\ref{eq_transmission_ij}).
On the other hand, ``bendings" of the paths outside the ring are connected in series and provide difference in the amplitudes of the transmission. 
We can thus write $T_{ij}\propto [a_0+\cos\frac{\omega}{2}\cos(\phi+\delta)]$ for transmissions across the ring.

The general formula for the four-terminal resistance at absolute zero on this notation is given as eq.(8) in Ref.\citen{buettiker}. In the present non-local measurement, it reads
\begin{equation}
R_{13,24}=\frac{1}{G_{\rm q}}\frac{T_{21}T_{43}-T_{23}T_{41}}{D},
\label{eq_4term_resistance}
\end{equation}
where $D\equiv G_{\rm q}^{-2}S(\alpha_{11}\alpha_{22}-\alpha_{12}\alpha_{21})$,
$S\equiv T_{12}+T_{14}+T_{32}+T_{34}$. And $\alpha_{ij}$ are sums of $T_{kl}$ and $T_{kl}T_{mn}$ products, of
which we do not show the tedious explicit forms. 
As can be guessed from the configuration shown in Fig.\ref{f1}(b), $T_{13}$ and $T_{24}$ are more than an order of magnitude larger than other coefficients and get much weaker effects from the interference on the ring.
This leads to an approximation $\alpha_{11}\alpha_{22}-\alpha_{12}\alpha_{21}\sim T_{13}T_{24}$.
Because $S$ consists of transmission coefficients across the ring, 
the effect of interference on the denominator $D$ can be summarized as $D\propto [a_0+\cos\frac{\omega}{2}\cos(\phi+\delta)]$.
On the other hand, terms in the numerator has a common factor $[a_0+\cos\frac{\omega}{2}\cos(\phi+\delta)]^2$ and we
eventually get
\begin{equation}
R_{13,24}\propto \left[a_0+\cos\frac{\omega}{2}\cos(\phi+\delta)\right].
\label{eq_dim_analysis}
\end{equation}

\begin{figure}
\includegraphics[width=\linewidth]{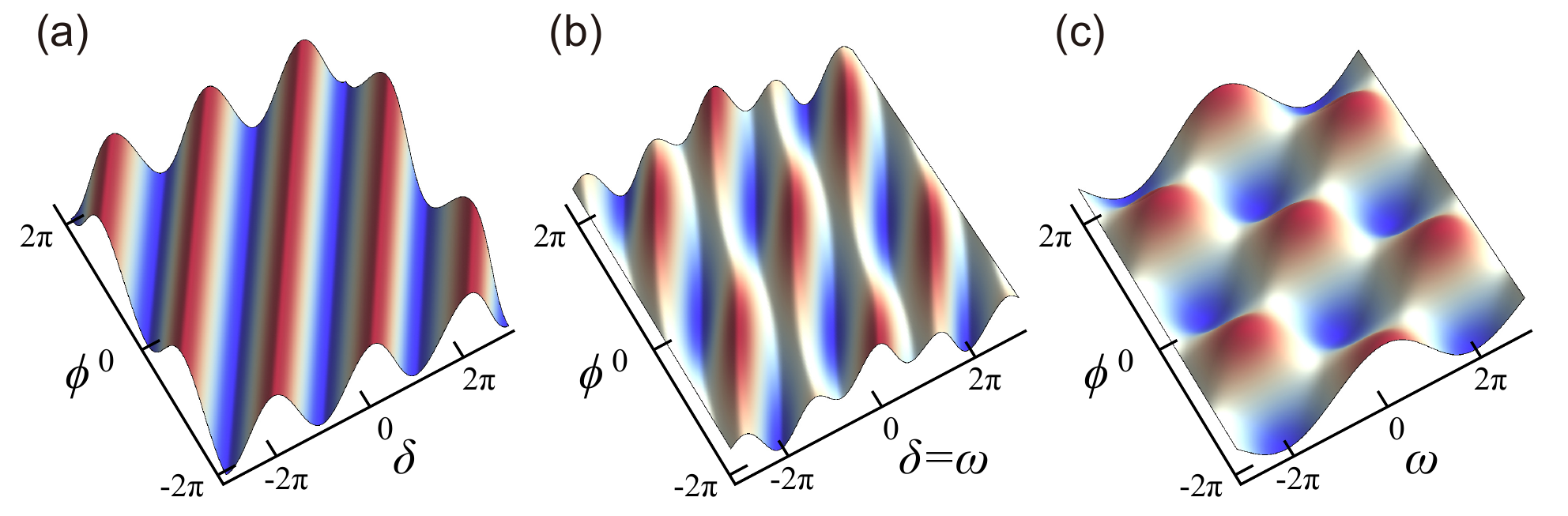}
\caption{(a) Interference pattern of four terminal non-local resistance $R_{13,24}$ calculated from eq.(\ref{eq_dim_analysis}). 
The form of RHS is plotted against the plane of the AB phase $\phi$ and the electro-statically shifted kinetic phase $\delta$. The phase shift in spin-space $\omega$ is kept constant (=0). (b) The same quantity as that in (a) but $\omega$ is proportional (in this plot equal) to $\delta$.
(c) $\delta$ is kept to 0 and the same is plotted versus $\omega-\phi$ plane.}
\label{f3}
\end{figure}

Figure \ref{f3} shows the interference patterns for (a) $\omega=0$, (b) $\omega=\delta$, (c) $\delta=0$.
It is well known that $\delta$ can be tuned with Schottky gate voltages\cite{kobayashi} and a pattern similar to Fig.\ref{f3}(a) was actually
reported in refs.\citen{kobayashi,takada}.
Then, if the gate voltage influences the phase $\omega$ simultaneously, a cross-hatching pattern like Fig.\ref{f3}(b) should appear and with increasing the sensitivity in $\omega$, the pattern should change to a checker board pattern in Fig.\ref{f3}(c).
The above analysis can then be summarized that
simple linear ``flow" of AB oscillation with a gate voltage on one arm is the sign of electrostatic modulation of kinetic phase 
and with overlapping of spin-interference, the angle of cross-hatching increases from 0 to 90$^\circ$.
Hence what we should look for is the behavior in Fig.\ref{f3}(b) or ultimately in Fig.\ref{f3}(c).

\section{Results}

We measured 4-terminal non-local resistance ($R_{13,24}$) as a function of the side gate voltage $V_{\rm R}$ with different AB phase $\phi$ and QPC conductance $G_{\rm QPC}$. AB phase $\phi$ is defined as $\phi=(B_{0} - B)/\phi_{0}$ where $B_{0}$ = $-$30~mT and $\phi_{0}$ is AB oscillation period calculated from the ring area (Fig.\ref{f1}(a)).
In order to obtain sufficient signal-to-noise ratio in the non-local measurement, a comparatively high bias voltage of 2~mV for the emitter QPC is required. This is reasonable considering low transmission coefficients over the AB ring.

\begin{figure}
\includegraphics[width=\linewidth]{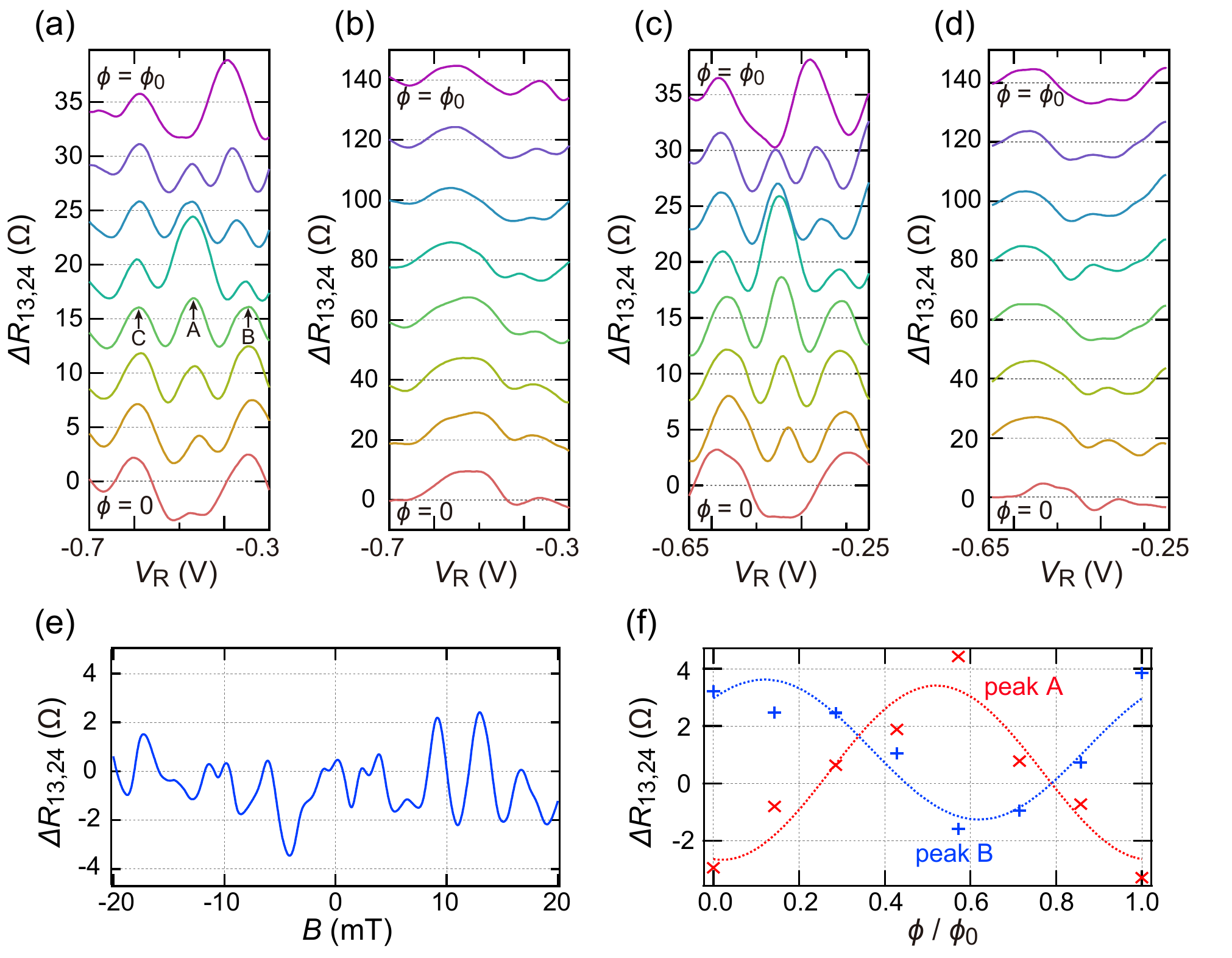}
\caption{(a-d) $\Delta R_{13,24}$ as a function of $V_{\rm R}$ with $\phi = 0$  (bottom) to $\phi = \phi_{0}$ (top). (a) and (c) corresponds to $G_{\rm QPC} = 1.0$ with $5 \Omega$/step offsets while (b) and (d) to $G_{\rm QPC} = 1.8$ with $20 \Omega$/step offsets. The bias voltage $V_{\rm b}$ was $-2.0$~mV (electron injection) for (a) and (c), and $+2.0$~mV (hole injection) for (b) and (d). (e) AB oscillation in four-terminal non-local resistance $\Delta R_{13,24}$. A background with period larger than 20 mT was subtracted. (f) Markers are plots of peak height vs AB phase $\phi / \phi_{0}$ for A and B peak illustrated in (a). Dashed lines are fitting sinusoidal curves calculated from least square method. }
\label{f4a}
\end{figure}

\begin{figure}
\includegraphics[width=\linewidth]{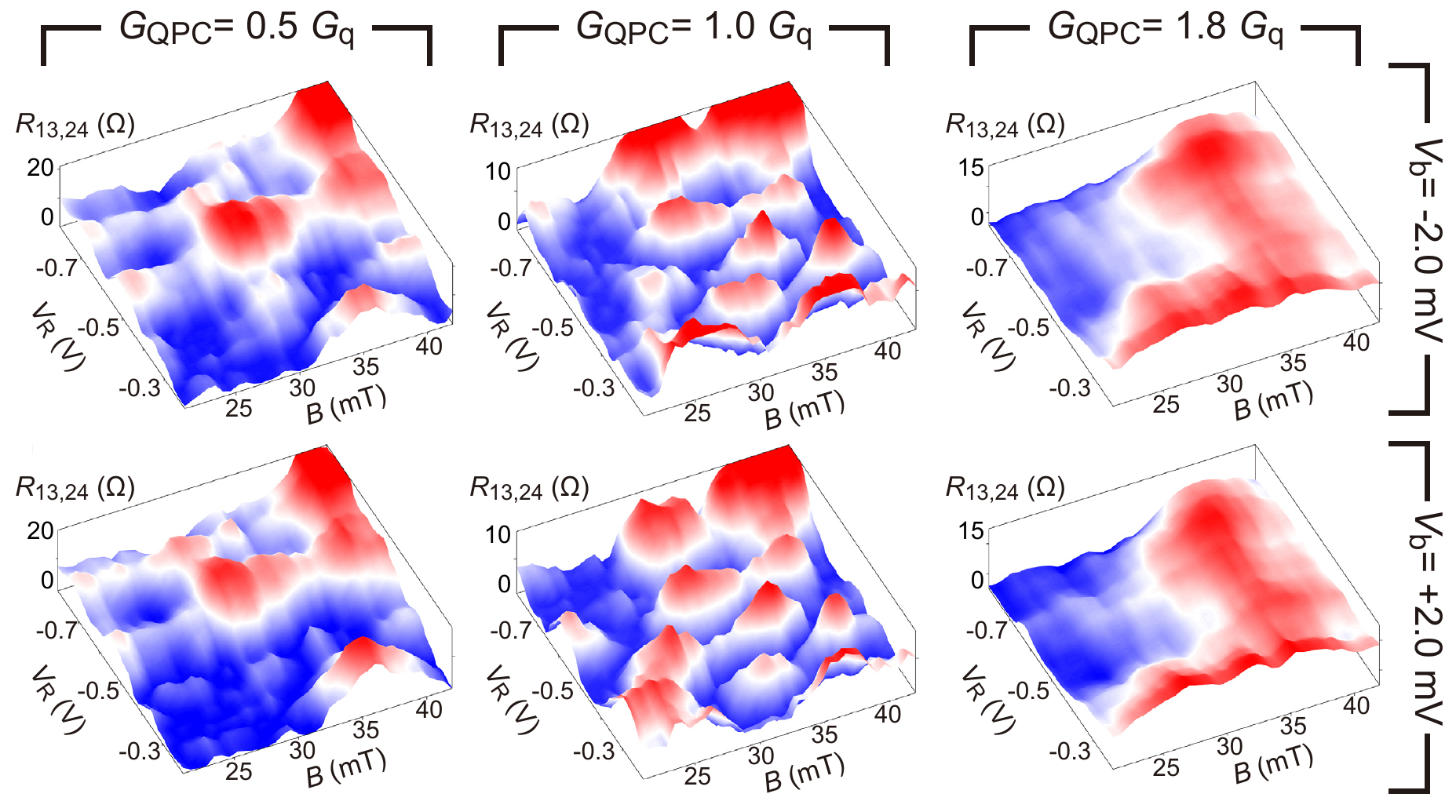}
\caption{Color plots of $R_{13,24}$ on the plane of $V_{\rm R}$ and magnetic field $B$. The columns from left to right are for $G_{\rm QPC}=0.5G_{\rm q}$, $1.0G_{\rm q}$, $1.8G_{\rm q}$ respectively. The upper row is for the QPC bias voltage $V_{\rm b}=-$2~mV (electron injection) and the lower for $V_{\rm b}=+$2~mV (hole injection).}
\label{f4b}
\end{figure}

In the case of $G_{\rm QPC}=1.8 G_{\rm q}$, Fig.\ref{f4a} (b, d) exhibit some peaks, though they were almost stable against $\phi$. On the other hand, by adjusting $G_{\rm QPC}$ to $1.0 G_{\rm q}$ peaks started to oscillate with a period of AB oscillation $\phi_{0}$ (Fig.\ref{f4a}(a, c)). 
Actually we observed a non-local AB oscillation when $G_{\rm QPC}=1.0 G_{\rm q}$ (Fig.\ref{f4a}(e)), which is asymmetric to the zero-magnetic field ensuring 4-terminal measurement is realized.
There appeared in Fig.\ref{f4a}(a) 3 peaks around $V_{\rm R}=-0.48$~V (peak A), $V_{\rm R}=-0.38$~V (peak B) and $V_{\rm R}=-0.60$~V (peak C). The height of peak A and B can be fitted by sinusoidal curve with phase difference $\sim \pi$ as illustrated in Fig.\ref{f4a}(f). This phase shift matches with the cross-hatching pattern in Fig.\ref{f3}(c). In fact Fig.\ref{f4b} shows such plots, where in the case of $1.0 G_{\rm q}$ we clearly observe oscillations depending on $V_{\rm R}$ and $\phi$. The patterns were stable by changing from electron to hole injection, which agrees well with the Landauer-B\"uttiker formalism\cite{buettiker} and thus guaranteed reproducibility of the measurement. Comparing to the case of $G_{\rm QPC}=1.0 G_{\rm q}$, Fig.\ref{f4b} for $G_{\rm QPC}=1.8 G_{\rm q}$ did not exhibit any cross-hatching pattern. However, for $G_{\rm QPC}\sim 0.5G_{\rm q}$, actually some weak trace of cross hatching pattern is observed. These results can be explained as following.
As noted before, high spin polarization at the emitter can be obtained when the QPC conductance is on the plateau of 0.5$G_{\rm q}$ and on the plateau of 1.0$G_{\rm q}$.
However as in Ref.\citen{kim}, the spin polarization on the 0.5$G_{\rm q}$ plateau is strongly reduced with increasing the bias voltage and around 2~mV the spin polarization is very small.
On the other hand, the spin polarization is almost constant or even increases with the bias voltage on 1.0$G_{\rm q}$ plateau. We can thus interpret the result in Fig.\ref{f4b} that the cross-hatching spin interference pattern appears when large pure spin current is injected into the device, reflecting the spin polarization of injected electrons.

\section{Discussion}
The results shown in Fig.\ref{f4b} were not predicted  in the theory in Ref.\citen{aharony}, in which the AB oscillation amplitude does not directly reflect the initial degree of spin polarization. This comes from the special symmetric device setup in Ref.\citen{aharony}.
There is also a quantitative difficulty in the interpretation of the above result.
Assuming Rashba SOI as the origin of spin phase shift, we can estimate the oscillation period in the gate voltage as\cite{nitta2}
\begin{equation}
\Delta V_{\rm R} = 2\pi/\omega = \hbar^2/2m^*\alpha L \sim 50~{\rm V},
\label{eq_rashba_estimation}
\end{equation}
which is much larger than the observed result $\sim$ 100~mV. 
Rather the observed value matches well with the Coulomb peak period of strongly-coupled states.
This urges us to look for a different origin of spin rotation around the quantum dot.

\begin{figure}
\includegraphics[width=\linewidth]{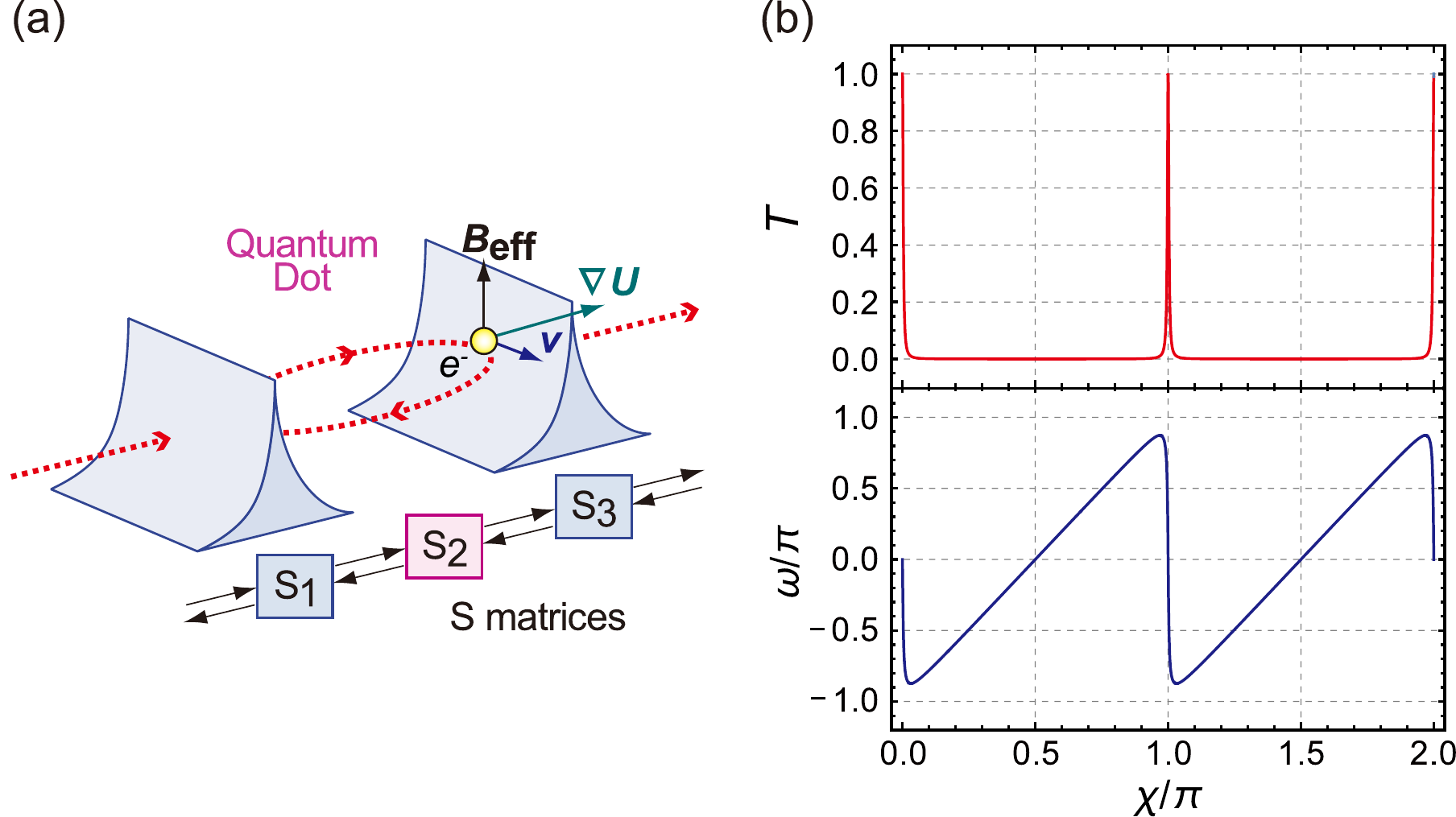}
\caption{(a) Simple resonator model of a quantum dot with a ellipsoidal orbit inside. The potential barrier gradient $\nabla U$ and the transverse velocity $v$ produce a effective magnetic field $B_{\rm eff}$ rotating the spin. These are hence further abstracted as the S-matrix circuit shown in the bottom. (b) Transmission coefficient $T$ and accumulated spin rotation $\omega /\pi$ as a function of spin phase $\chi$ at the potential wall in the model in (a). The parameter is $r_{0} = 0.99$.}
\label{f5}
\end{figure}

Here we point out an efficient spin rotation mechanism in a quantum resonator like a quantum dot.
A quantum dot resonator can be modeled with two potential barriers as in Fig.\ref{f5}(a).
We assume it has a classically ellipsoidal orbit as illustrated in the figure.
It is well known that confinement potentials $U$ for semiconductor quantum dots are well approximated as harmonic potentials and thus have large potential gradients $\bvec{\nabla}U$.
Through the SOI
\begin{equation}
\mathcal{H}_{\mathrm{SOI}}\sim \bvec{\sigma}\cdot \left(\bvec{p}\times \bvec{\nabla}U \right),
\end{equation}
this gives an effective magnetic field perpendicular to the conduction plane 
and consequent spin phase rotation $\chi$. 
With repetition of reflection, spin standing wave emerge with accumulated spin rotation $\omega$. 
The scatterings at the barriers can mathematically be expressed with S-matrices as
\begin{equation}
S_{1}=S_{3}=\pmatrix{
r_{0}e^{+i\chi \sigma_{z}} & t_{0} \cr
t_{0} & r_{0}e^{+i\chi \sigma_{z}} \cr
},\quad S_{2}=\pmatrix{
0 & 1 \cr
1 & 0 \cr
},
\label{eq_s_matrices}
\end{equation}
where $\sigma_{z}$ is the $z$ component of Pauli matrix. After calculating the composition of S-matrices, we get the following expression for the total transmission,

\begin{equation}
t_{\rm total} = \pmatrix{
t_{0}^2\ \Big/\left(1-r_{0}^2 e^{+2i\chi} \right) & 0 \cr
0 & t_{0}^2\ \Big/\left(1-r_{0}^2 e^{-2i\chi}\right) \cr
} \equiv \pmatrix{
t_{+} & 0 \cr
0 & t_{-} \cr
}.
\label{eq_t_total}
\end{equation}

Figure \ref{f5}(b) are the transmission coefficient $T = \big| t_{+} \big|^2 + \big| t_{-} \big|^2$ and the accumulated spin rotation $\omega = \arctan (t_{-}/t_{+})$ respectively, which are calculated from eq.(\ref{eq_t_total}) and the model circuit shown in Fig.\ref{f5}(a).
As expected the spin rotates by $2\pi$ within a peak-to-peak interval.
Note that the resonances appearing in Fig.\ref{f5}(b) are pure ``spin resonance" though in the actual system, spin and orbital are entangled through SOI and a $\pi$ rotation between adjacent Coulomb peaks is expected.
This perfectly agrees with the present observation because the phase shift through a quantum dot is dominated by the strongly-coupled states\cite{aikawa} and the cross-hatch period should be the one for strongly-coupled states observed in Fig.\ref{f2}(b).

We should comment on why high spin current is essential for a clear cross hatched pattern. The total spin phase depends on the initial spin state as calculated in eq.(\ref{eq_t_total}). Thus spin phase of unpolarized electrons, that is electrons with random polarization, has random distribution resulting in disappearance of the interference pattern. Though this effect had made it difficult for us to observe the spin interference, now we can make use of it conversely to check the operation of the quantum dot spin rotators by virtue of the QPC spin injector.
Fig.\ref{f4b} clearly exhibits the disappearance of the interference pattern when unpolarized electrons were injected ($G_{\rm QPC}\sim 1.8 G_{\rm q}$). It guarantees that $V_{\rm R}$ did modulate the spin phase via the quantum dot spin rotators.

\section{Conclusion}

In summary, we have fabricated a spin interference device with a spin injector, which utilizes a QPC and Rashba SOI.
A clear spin interference signal is obtained only when highly spin-polarized electrons or holes are injected.
From the observation we have presented a reasonable model of a resonator with a SOI and shown that such a resonator can work as an efficient spin rotator controllable with gate voltage.

\ack

We thank A. Aharony and O. Entin-Wohlman for valuable discussion.
This work was supported by Grant-in-Aid for Scientific Research on Innovative Area, ``Nano Spin Conversion Science" (Grant No.26103003), also by Grant No.25247051 and by Special Coordination Funds for Promoting Science and Technology. 
Iwasaki was also supported by Japan Society for the Promotion of Science through Program for Leading Graduate Schools (MERIT).

\end{document}